\documentstyle[pra,aps,epsf,twocolumn]{revtex}

\def\IncludeEpsImg#1#2#3#4{\renewcommand{\epsfsize}[2]{#3##1}{\epsfbox{#4}}}

\def\ipcx#1#2#3{\raisebox{0mm}{\hspace{#1mm}\raisebox{#2mm}{\IncludeEpsImg{5.75mm}{3.76mm}{.300}{#3.eps}}}}

\def\cm#1{}

\begin{document}
\sloppy
\title{Perturbation Theory for Particle in a Box}

\author{H.\ Kleinert,\thanks{E-mail: kleinert@physik.fu-berlin.de}
     A.\ Chervyakov\thanks{On leave from LCTA, JINR, Dubna, Russia,
                           E-mail: chervyak@physik.fu-berlin.de   },
     and B.\ Hamprecht \\
          Institut f\"ur Theoretische Physik\\
          Freie Universit\"at Berlin\\
           Arnimallee14, D-14195 Berlin}
\maketitle
\begin{abstract}
Recently developed strong-coupling theory open up the possibility
of treating quantum-mechanical systems with hard-wall potentials
via perturbation theory. To test the
power of this theory we study here
the exactly solvable quantum mechanics of a
point particle in a one-dimensional box. Introducing
an auxiliary harmonic mass term $m$,
the ground-state energy $E^{(0)}$ can  be expanded perturbatively
in powers
of $1/md$, where $d$ is
the box size.
The removal of the infrared cutoff $m$
requires the resummation of the series at infinitely strong coupling.
We show that strong-coupling theory
yields a fast-convergent sequence of approximations to
the well-known quantum-mechanical energy $E^{(0)}= \pi ^2/2d^2$.
 \end{abstract}

{\bf 1}. Variational perturbation theory
\cite{3}
permits us to convert
divergent
weak-coupling expansions into
convergent strong-coupling expansions.
In particular, a constant strong-coupling limit
of a function
can be evaluated from its weak-coupling expansion
with any desired accuracy.
As an important application, this has led
to a novel way of calculating  critical exponents
without using
the renormalization group \cite{kl}.

Given this theory, new classes of physical
systems become accessible to perturbation theory.
For instance, the important problem
a
the pressure exerted by a stack of
membranes
upon the enclosing walls \cite{1} has now become calculable
analytically with the help of perturbation theory.
For a single membrane, this has already be done successfully
\cite{2,2B}.
Realistic physical problems have usually the disadvantage
that the maximally accessible order of perturbation theory
is quite limited. If we want to gain a
better understanding of the
convergence of the successive approximations
as the order goes to infinity
it is useful to study a system where the
result is known exactly.
This will be done in the present note
for a
quantum-mechanical point particle in a
one-dimensional box. The ground state energy of this system
is
known exactly, $E^{(0)}=  \pi ^2/2d^2$ (in natural units),
where $d$ is size the of the box.
We shall demonstrate how this result
is found via
strong-coupling theory
from a perturbation expansion,
thus illustrating
the reliability
of the earlier membrane
calculations \cite{2,2B}.

{\bf 2}. The partition function of a particle in a box is given by the
euclidean path integral (always in natural units)
\begin{equation}
 Z = \int  {\cal D} u(t) e^{\frac{1}{2} \int dt ( \partial u)^2}
,
\label{1}\end{equation}
where the particle coordinate $u(t)$ is restricted to the interval
$-d/2 \leq u(t) \leq d/2$. Since such a hard-wall
restriction
is difficult to treat analytically in the path integral (\ref{1}),
we make the hard-walls soft by adding to the euclidean action $E$ in the
exponent of (\ref{1})
a potential term
diverging near the walls.
Thus we consider the auxiliary euclidean action
\begin{equation}
  E = \frac{1}{2} \int dt \left\{  [\partial u(t)]^2 +
    V(u(t))\right\} ,
\label{2}\end{equation}
  where $V(u)$ is given by
\begin{equation}
 \!\!\!\!  V (u) = \frac{\omega ^2}2 \left(\frac{d}{ \pi }
\tan \frac{ \pi u}{d}\right)^2=
  \frac{ \omega ^2}{2}\left(u^2 + \frac{2}{3}\,gu^4+\dots\right).
\label{3}\end{equation}
On the right-hand side we have introduced a parameter  $g \equiv  \pi ^2 /
d^2$.

\begin{figure}[tbhp]
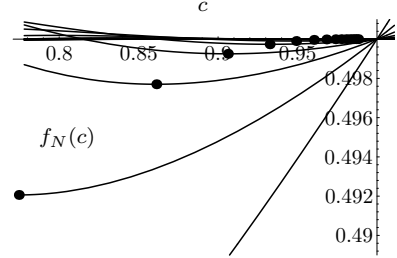

\input shmin.tps
\caption[]{Variational functions $f_N(c)$ up to $N=16$
are shown together with their minima whose $y$-coordinates approach rapidly the
correct
limiting value $1/2$.
}
\label{@min}\end{figure}

{\bf 3}. The expansion of the potential in powers of $g$
can now be treated perturbatively,
leading to an expansion of $Z$
around the harmonic part of the partition function,
in which
the integrations over $u(t)$
run over the entire
$u$-axis and yield
\begin{equation}
  Z_ \omega  = e^{-(1/2) {\rm Tr} \log (\partial ^2 +  \omega ^2)}.
\label{4}\end{equation}
For $L\rightarrow \infty$,
the exponent gives a
free energy density $f=-L^{-1}\log Z$
equal to
 the ground state energy
 of the harmonic  oscillator
\begin{equation}
f_0=\frac{ \omega }{2}.
\label{@}\end{equation}
The treatment of the interaction terms can be organized in powers of $g$,
and give rise to an expansion of the free energy
with the generic form
\begin{equation}
f=f_0+ \omega \sum _{k=1}^\infty a_k \left(\frac{g}{ \omega }\right)^k.
\label{@genfor}\end{equation}
The calculation of the coefficients
$a_k$ in this expansion
proceeds as follows.
First we expand
the potential in (\ref{2})
to identify the power series for
the interaction energy
\begin{eqnarray}
  E^{\rm int}  &= & \frac{ \omega ^2}{2} \int dt \left\{ g \varepsilon_4 u^4 +
     g^2 \varepsilon_6 u^6 + g^3 \varepsilon_8 u^8 + \dots \right\} \nonumber
\\
      & = &  \frac{ \omega ^2}{2} \sum_{k=1}^{\infty} \int dt \, g^k
         \varepsilon_{2k + 2} [u^2  (t)]^{k+1} \,,
\label{5}\end{eqnarray}
 with coefficients
\begin{eqnarray}
&& \varepsilon_4 = \frac{2}{3}, ~~\varepsilon_6 = \frac{17}{45},~~
\varepsilon_8 = \frac{62}{315},
{}~~ \varepsilon_{10} = \frac{1382}{14175},\nonumber\\
&&~~\varepsilon_{12} = \frac{21844}{467775},
{}~~ \varepsilon_{14} = \frac{929569}{42567525},
{}~~ \varepsilon_{16} = \frac{6404582}{638512875},\nonumber\\
&&~~\varepsilon_{18} = \frac{443861162}{97692469875},
{}~~ \varepsilon_{20} = \frac{18888466084}{9280784638125}, \nonumber\\
&&~~ \varepsilon_{22} = \frac{113927491862}{126109485376875},
{}~~\varepsilon_{24} = \frac{58870668456604}{147926426347074375},\nonumber\\
&&~~ \varepsilon_{26} = \frac{8374643517010684}{48076088562799171875},
\nonumber\\
&&~~ \varepsilon_{28} =
\frac{689005380505609448}{9086380738369043484375},\nonumber\\
&&~~\varepsilon_{30} =
\frac{129848163681107301953}{3952575621190533915703125},\nonumber\\
&&~~ \varepsilon_{32} =
\frac{1736640792209901647222}{122529844256906551386796875},
\nonumber\\
&&~~ \varepsilon_{34} =
\frac{418781231495293038913922}{68739242628124575327993046875},
 \dots\,.
\label{6}\end{eqnarray}
The interaction terms $\int dt \,[u^2  (t)]^{k+1}$ and their products
are expanded
according to Wick's
rule into sums of
products of
Wick contractions representing
harmonic two-point correlation functions
\begin{equation}
 \langle u(t_1) u(t_2) \rangle = \int \frac{dk}{2 \pi }
     \, \frac{e^{ik(t_1 -t_2) }}{k^2 +  \omega ^2} =\frac{e^{- \omega
|t_1-t_2|}}{2 \omega }.
\label{7}\end{equation}
Associated local expectation values are
$\langle u^2 \rangle  =
{1}/{2 \omega }$, and
\begin{eqnarray}
     \langle u \partial u\rangle & = & \int
          \frac{dk}{2 \pi } \frac{k}{k^2 +  \omega ^2} = 0 \nonumber\\
   \langle \partial u \partial u\rangle & = & \int
    \frac{dk}{2 \pi }  \frac{k^2}{k^2 +  \omega ^2} = - \frac{ \omega }{2},
 \label{8}\end{eqnarray}
where the last integral is calculated using
dimensional regularization
in which $\int dk\,k^ \alpha =0$ for all $ \alpha $.
The Wick contractions are organized with the help of Feynman diagrams.
Only the connected diagrams contribute to the free energy density.
The graphical expansion of
free energy
up to four loops is
\begin{eqnarray}
&&f  =  \frac{ \omega }{2}+ \left(\frac{ \omega ^2}{2}\right)
         \left\{ g   \varepsilon_4  \,  3\,   \ipcx{0}{-.4}{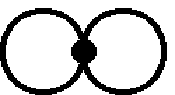}
     + g^2   \varepsilon_6   15 \ipcx{1}{-1.8}{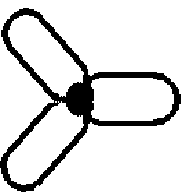}+
   g ^3   \varepsilon_8   105 \ipcx{1}{-1.5}{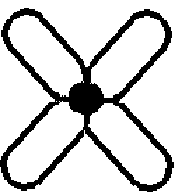}\right\} \nonumber \\&& -
\frac{1}{2!}
   \left(\frac{ \omega ^2}{2}\right)^2 
\left\{  g^2  \varepsilon_4^2 \left[ 72  \ipcx{1}{-.1}{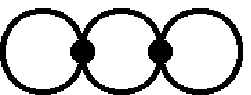}~\,+
    24  \ipcx{1}{-.5}{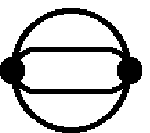}\right] \right. \nonumber \\
 & & ~~~~~~~~~~~~~~~~\left.+ g^3\,  2  \varepsilon_4  \varepsilon_6 \left[
540  \ipcx{1}{-2.}{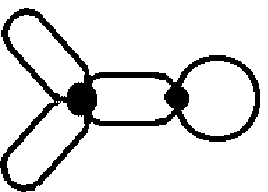}~~    + 360 \ipcx{1}{-.8}{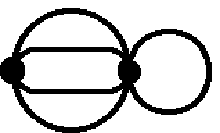}~\right] \right\}
\nonumber \\
 & & + \frac{1}{3!} \left(\frac{ \omega ^2}{2} \right)^3
     g^3  \varepsilon_4^3  \left\{ 2592 \ipcx{1}{-0}{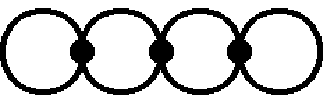}~~~ +
  1728 \ipcx{1}{-1.5}{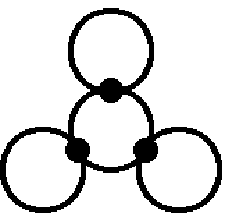}
\right.
\nonumber \\ & & ~~~~~~~~~~ ~~~~~~~~~~~~~
\left.
 + 3456  \ipcx{1}{-.5}{f4-5} +
1728  \ipcx{1}{-.5}{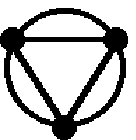}
     \right\}
{}.
\label{9}\end{eqnarray}
Note different numbers of loops contribute to the terms
of  order $g^n$.
The calculation of the diagrams in Eq.~(\ref{9}) is simplified by
the factorization property: If a diagram consists of two subdiagrams
touching each other at a single vertex, the associated
Feynman integral factorizes into
those of the subdiagrams. In each diagram, the last $t$-integral yields
 an overall factor $L$, due to translational invariance along
the $t$-axis, the others produce a factor $1/\omega $.
 Using the explicit expression
(\ref{8}) for the lines in the diagrams,
we find the following velues for the Feynman integrals:
%
%
\begin{eqnarray}
\raisebox{-.5mm}{\ipcx{0}{.2}{f3-2}}~~~~
& = & L \, \frac{1}{16  \omega ^5},
{}~~~~\raisebox{-1mm}{\ipcx{0}{0.5}{f4-3}}~~~~
= L  \, \frac{1}{64 \omega ^8},
      \nonumber \\
\raisebox{-.2mm}{\ipcx{0}{-.1}{f3-1}}~~~~
& = & L \, \frac{1}{32 \omega ^5},
{}~~~~~\raisebox{.2mm}{\ipcx{0}{-.7}{f4-4}}~~~
 = L \, \frac{3}{128  \omega ^8},
    \nonumber \\
\raisebox{1mm}{\ipcx{0}{-2.5}{f4-6}}~~~~
 & = & L \, \frac{1}{32 \omega ^6} ,
{}~~~~~\;\raisebox{-.2mm}{\ipcx{0}{-2}{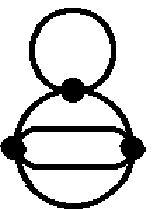}}~~~\!
= L \, \frac{5}{8 \cdot 64 \cdot
        \omega ^8} , \\
\ipcx{0}{-1}{f4-5}~~~~
& = & L \, \frac{1}{32  \omega ^6} ,
{}~~~~~~\ipcx{0}{-1}{f4-1}~~
= L \,
   \frac{3}{8 \cdot 64 \cdot  \omega ^8} \nonumber
\label{10}\end{eqnarray}
Adding all contributions in (\ref{9}),
we obtain  up to the order $g^3$:
\begin{eqnarray}
   f_3 & = &  \omega \left\{ \frac{1}{2} + \frac{3}{8} \varepsilon_4
      \left(\frac{g}{ \omega }\right) +
     \left[ \frac{15}{16} \varepsilon_6 - \frac{21}{32} \varepsilon_4^2 \right]
      \left(\frac{g}{ \omega }\right)^2 +\right. \nonumber \\
 && \left. + \left[ \frac{105}{32} \varepsilon_8 - \frac{45}{8} \varepsilon_4
\, \varepsilon_6 +
     \frac{333}{128} \varepsilon_4^3 \right] \left(\frac{g}{ \omega }\right)^3
   \right\},
\label{11}\end{eqnarray}
which has the generic form
(\ref{@genfor}).
We can go to higher orders
by extending the
Bender-Wu recursion relation for the ground-state
energy of the quartic anharmonic oscillator
as follows:
\begin{eqnarray}
&& 2 j C_{n,j}  = (j+1)(2j+1) C_{n,j} - \nonumber \\
&&    \frac{1}{2} \sum_{k=1}^{n} (-1)^k \varepsilon_{2k+2} C_{n-k},j-k -1
\nonumber \\
&& - \sum_{k=1}^{n-1} C_{k,1} C_{n-k}, j,~~~1 \leq j \leq 2n,\nonumber\\
&& C_{0,0}=1 , ~~C_{n,j}  = 0 ~~~~(n \geq 1, j < 1).
\label{13}\end{eqnarray}
After solving these recursion relations,
the coefficients
$a_k$ in (\ref{@genfor}) are given by
$a_k = (-1)^ {k+1} C_{k,1}$.
For brevity, we list here
the first  sixteen
expansion coefficients
for $f$,
calculated with the help of the algebra program
REDUCE:
\begin{eqnarray}
&&\,a_0 = \frac{1}{2},\, a_1 = \frac{1}{4},\, a_2 = \frac{1}{16},
\,a_3 = 0,\,a_4 = -\frac{1}{256}, \,a_5 = 0,\nonumber \\
&&\,a_6 = \frac{1}{2048},\, a_7 = 0,\,a_8 = -\frac{5}{65536},\,a_9 = 0,
\nonumber\\
&&\,a_{10} = \frac{7}{524288},\, a_{11} = 0,\,a_{12} = -\frac{21}{8388608},
\,a_{13} = 0,\,\nonumber \\&&\, a_{14} = \frac{33}{67108864},
\,a_{15} = 0,\,a_{16} = -\frac{429}{4294967296}, \dots~.
\label{14}\end{eqnarray}
%
{\bf 4}. We are now ready to
calculate successive
strong-coupling approximations to the function $g(g)$
and suty
the convergence behavior as the order grows large.
According to the general theory in \cite{kl,3}, the
$N$th order approximation
to the strong-coupling limit $f^*$ of the series $f(g)$
is found
by replacing,  in the truncated series after the $N$th term $f_N(g)$,
the frequency $ \omega $ by
the identical expression $\sqrt{ \Omega ^2+ gr}$,
where $r\equiv ( \omega ^2- \Omega ^2)/g$ which is, however,
treated  for a moment as an independent
variable, whereas $ \Omega $ is a dummy parameter.
Then the square root is expanded binomially in powers of $g$,
and $f_N(g)$ is reexpanded up to
order $g^N$. After that, $r$ is replaced  by its proper value.
In this way we obtain a function
$f_N(g, \Omega )$
which depends on $ \Omega $, which thus becomes a variational
parameter. The best approximation is obtained by extremizing
$f_N(g, \Omega )$ with respect to $ \omega $.
The strong-coupling limit  is obtained by
taking $g$ to infinity which is equivalent to setting $ \omega =0$.
In this limit, the optimal $ \Omega $ will grow
proportionally with $g$, so that $g/ \Omega =1/c$ is finite,
and the variational expression
$f_N(g, \Omega )$ becomes a function of $f_N(c)$.
For $ \omega =0$,
the above reexpansion amounts simply to replacing
each power $ \omega^n $ in each  expansion terms of
$f_N(g)$ by
the binomial expansion
of $ (1-1)^{-n/2}$ truncated after the $(N-n)$th term.
The first nine variational functions
 $f_N(c)$   are listed in Table~1.
\begin{table}[tbhp]
\caption[]{First eight variational functions $f_N(c)$.
}
\scriptsize{
\begin{tabular}{l}
$f_2(c)= {1\over 4} + {1\over {16\,c}} + {{3\,c}\over {16}} $\\
$f_3(c)=  {1\over 4} + {3\over {32\,c}} + {{5\,c}\over {32}}$\\
$f_4(c)=
  {1\over 4} - {1\over {256\,{c^3}}} + {{15}\over {128\,c}} +
   {{35\,c}\over {256}}$\\
$f_5(c)={1\over 4} - {5\over {512\,{c^3}}} +
   {{35}\over {256\,c}} + {{63\,c}\over {512}}$\\
$f_6(c)=
  {1\over 4} + {1\over {2048\,{c^5}}} - {{35}\over {2048\,{c^3}}} +
   {{315}\over {2048\,c}} + {{231\,c}\over {2048}}$\\
$f_7(c)=
  {1\over 4} + {7\over {4096\,{c^5}}} - {{105}\over {4096\,{c^3}}} +
   {{693}\over {4096\,c}} + {{429\,c}\over {4096}}$\\
$f_8(c)=
  {1\over 4} - {5\over {65536\,{c^7}}} + {{63}\over {16384\,{c^5}}} -
   {{1155}\over {32768\,{c^3}}} + {{3003}\over {16384\,c}} +
   {{6435\,c}\over {65536}}$\\
$f_9(c)={1\over 4} - {{45}\over {131072\,{c^7}}} +
   {{231}\over {32768\,{c^5}}} - {{3003}\over {65536\,{c^3}}} +
   {{6435}\over {32768\,c}} + {{12155\,c}\over {131072}}$
\end{tabular}                                                 }
\label{@}\end{table}
The functions $f_N(c)$ are minimized
starting from $f_2(c)$ and searching
the minimum
of each successive $f_3(c)$, $f_3(c),\dots~$
nearest to the previous one.
The functions $f_N(c)$ together with their minima are
plotted in Fig.~\ref{@min}.
\noindent The minima lie at%
\begin{eqnarray}
 (N,f^{\rm min}_N)&=&
 ( 2,0.466506) ,~( 3,0.492061) ,~( 4,0.497701) ,\nonumber \\&&( 5,0.499253) ,
  ~( 6,0.499738) ,~( 7,0.499903) ,~\nonumber \\&&( 8,0.499963) ,~( 9,0.499985)
,
  ( 10,0.499994) ,\nonumber \\&&( 11,0.499998) ,~( 12,0.499999) ,~( 13,0.5000)
, \nonumber \\&&
  ( 14,0.50000) ,~( 15,0.50000) , ~( 16,0.5000).
\label{@}\end{eqnarray}
They converge exponentially fast against the known result $1/2$, as shown in
Fig.~\ref{@conv}.
\begin{figure}[tbhp]
{}~\\
\input expon.tps
\caption[]{}
\label{@conv}\end{figure}

{\bf 5}. The alert reader will have noted that
the expansion coefficients
(\ref{14})
possesses two special properties:
First, they
lack the factorial growth
at large orders which would be found for a single power $[u^2  (t)]^{k+1} $
of the interaction potential
\cite{5}.
The factorial growth
is canceled by the specific combination of the different powers in the
interaction
(\ref{5}), making the series
(\ref{@genfor}) convergent
inside a certain circle.
Still, since this circle  has a finite radius (the ratio test shows that it
is unity),
this convergent series cannot be evaluated
in
the limit
of large $g$ which we want to do,
so that
variational
strong-coupling theory is not superfluous.
However, there is a second remarkable property of
the coefficients (\ref{14}): They contain
an infinite number of zeros in the sequence of
coefficients for each odd number,
except for the first one.
We may take advantage of this
property by separating
off the irregular term $a_1g=g/4=\pi^2/4d^2$, setting $ \alpha =g^2/4 \omega
^2$,
and
rewriting
$f$ as
\begin{eqnarray}
&&     f =\frac{\pi^2}{4d^2}\left[ 1+\frac{1}{ \sqrt{ \alpha } } h( \alpha
)\right],~~~~h( \alpha )\equiv
  \sum_{n=0}^{N} 2^{2n+1}
       a_{2n} \alpha^n .
\label{15}\end{eqnarray}
Inserting the numbers (\ref{14}),
the expansion of $  h( \alpha )$ reads
\begin{eqnarray}
  h( \alpha )&=&
  \left. 1 + \frac{\alpha}{2}
       - \frac{\alpha^2}{8} + \frac{\alpha^3}{16} - \frac{5}{128} \alpha^4
       + \frac{7}{256} \alpha^5 \right. -\nonumber \\
&&~~~~~~~~ \left. - \frac{21}{1024} \alpha^6  + \frac{33}{2048} \alpha^7 -
    \frac{429}{32768} \alpha^8 + \dots \right..
\label{15ex}\end{eqnarray}
We now
realize
 that this
is
the bibomial power series expansion of
$    \sqrt{1 + \alpha}$.
Substituting this into (\ref{15}), we find
the exact ground state energy
for the euclidean action
(\ref{2})
\begin{equation}
  E^{(0)} = \frac{ \pi ^2}{4d^2}\left(1  +  \sqrt{1 + \frac{1}\alpha}
\right)=
 \frac{ \pi ^2}{4d^2} \left(1 +  \sqrt{1 + 4 \omega ^2 \frac{d^4}{ \pi ^4}}
\right).
\label{17}\end{equation}
Here we can go directly to the strong-coupling limit  $ \alpha \rightarrow
\infty$
corresponding to $d\rightarrow 0$,
where we recover the
 exact
  ground-state energy $E^{(0)}= \pi ^2/2d^2$.

{\bf 5}. The energy (\ref{17}) can of course be
obtained directly by solving
the Schr\"odinger equation
associated with the potential (\ref{5}),
\begin{eqnarray}
\frac{1}{2}\left\{
-\frac{\partial ^2}{\partial x^2}+
\left[
\frac{ \lambda (1- \lambda) }{\displaystyle\cos^2x} -1
\right]\right\} \psi(x)=\frac{d^2}{\pi^2}E  \psi(x),
\label{@posch}\end{eqnarray}
where we have replaced
$u\rightarrow d x/\pi$ and set  $ \omega ^2d^4/\pi^4\equiv   \lambda ( \lambda
-1)$,
so that
\begin{eqnarray}
&&
   \lambda = \frac{1}{2} \left(1 +  \sqrt{1 + 4 \omega ^2
         \, \frac{d^4}{ \pi ^4}}\right).
\label{18la}\end{eqnarray}
Equation (\ref{@posch})
is of the
P\"oschl-Teller type and has the ground state wave function \cite{6}
\begin{eqnarray}
&& \psi_0(x) ={\rm const}\times  \cos^ \lambda  x\,,
\label{18}\end{eqnarray}
with the eigenvalue $\pi^2E^{(0)}/d^2=( \lambda ^2-1)/2$,
which agrees of xourse with Eq.~(\ref{17}).

If we were to apply the variational procedure to the series
$h( \alpha)/ \sqrt{ \alpha} $ in $f$ of Eq.~(\ref{17}),
by replacing the factor $1/ \omega^{2n}$
contained in each power $ \alpha^n$ by
$ \Omega= \sqrt{ \Omega^2+r \alpha}$
and reexpanding now in powers of $ \alpha$ rather than $g$, we would find that
all
approximation $h_N(c)$
would posses  a minimum with  unit  value, such that the corresponding
extremal functions $f_N(c)$ yield the correct final energy
in {\em each\/} order $N$.

\cm{Given the function $\nu (z) = \frac{1}{4}  \left(1 +  \sqrt{1+z^{-1}}
\right)$ we derive the common kth term
\begin{eqnarray}
   \varepsilon_k & = & \frac{1}{2 \pi i} \oint_{c} dz\nu (z)z^{k-1}\nonumber\\
       & = & -\frac{1}{2\pi i} \int^{0}_{-1} dz(1+z^{-1})^{1/2}z^{k-1}
     =  - \frac{(-1)^{k}}{4\sqrt{\pi}}\,\frac{(k-\frac{3}{2})!}{k!},
\label{19}\end{eqnarray}
 where the contour of unit radius $C$ enclosed the origin.
The expansion (\ref{15}) therefore takes the form
\begin{equation}
  f^\infty \big/ g = -\frac{1}{4  \sqrt{ \pi } } \sum_{n=0}^{\infty}
     \frac{ \Gamma (n - \frac{1}{2} )}{ \Gamma (n+1)} (-\alpha)^{-n}.
\label{20}\end{equation}
This has the large $-\alpha$ limit:
$f^\infty\big/ g \rightarrow -  \Gamma \left(-\frac{1}{2}\right)\big /
4 \sqrt{ \pi }  = 1/2$ which is the same as for $f^N\big / g$,
 because the remainder $R_N(\alpha) = \sum_{n=N+1}^{\infty}  \varepsilon_n
 (-\alpha)^{-n}$ becomes very small to this limit. Indeed,
since $\varepsilon_n$ decreases as $\varepsilon_n \rightarrow (e/n)^{3/2}$,
$n \rightarrow \infty$, for any coupling $\alpha^* : |\varepsilon_n
(\alpha^*)^{-n}|
\leq K$ and $|\alpha| > \alpha^*$, we estimate
\begin{equation}
| R_N (\alpha) | \leq \sum_{N+1}^{\infty} K
\left(\frac{\alpha^*}{|\alpha|}\right)^n
   = K \frac{(\alpha^*/|\alpha|)}{1 -\alpha^*/|\alpha|},
\label{21}\end{equation}
which explain completely the above over-convergence.}


\begin{thebibliography}{11}

\bibitem{3}
H.\ Kleinert,  {\em Path Integrals in Quantum Mechanics, Statistics
   and Polymer Physics\/}, World Scientific, Singapore, 1995.

\bibitem{kl}
H. Kleinert,
Phys. Rev. D {\bf 57}, 2264 (1998) (APS E-Print aps1997jun25\_001); Adendum:
ibid. D {\bf 58}, 1077 (1998) (cond-mat/9803268);
{\em Strong-Coupling $\phi^4$-Theory in $4- \epsilon$ Dimensions and Critical
Exponents\/},
(cond-mat/9801167);
{\em Seven Loop Critical Exponents from Strong-Coupling
$\phi^4$-Theory in Three Dimensions\/},  Berlin preprint 1998
(hep-th/9812197).

\bibitem{1}
W.\ Helfrich, Z.\ Naturforsch.\ A {\bf 33}, 305 (1978), W.\ Helfrich,
 R.M.\ Servuss, Nuovo Cim.\ D {\bf 3}, 137 (1984);\\
W.\ Janke, H.\ Kleinert, Phys.\ Lett.\ {\bf 58}, 144 (1987),
W.\ Janke, H.\ Kleinert and H.\ Meinhardt, Phys.\ Lett.\
 B {\bf 217}, 525 (1989);\\
G.\ Gompper and D.M.\ Kroll, Europhys.\ Lett.\ {\bf 9}, 58 (1989),
 R.R.\ Netz and  R.\ Lipowski, Europhys.\ Lett.\ {\bf 29}. 345 (1995),
F.\ David, J.\ de Phys.\ {\bf 51}, C7-115 (1990).
\bibitem{2}
H.\ Kleinert, {\em Pressure of Membrane between Walls, \/}
 Berlin preprint, 1998 (cond-mat/9811308).
\bibitem{2B}
M. Bachmann,
H. Kleinert, and
A. Pelster,
{\em Strong-Coupling Calculation of the Fluctuation
Pressure of a Membrane Between Walls\/},
Berlin preprint 1999 (cond-mat/9905397).
\bibitem{5}
  C.M.\ Bender and T.T.\ Wu, Phys.\ Rev.\ {\bf 184}, 1231 (1969);
  \\
 Phys.\ Rev.\ D {\bf 7}, 1620 (1973).

\bibitem{6}
  S.\ Fl\"ugge,  {\em Practical Quantum Mechanics I\/},
   Springer-Verlag, Berlin-Heidelberg-New York, 1971.


\bibitem{4}
 H.\ Kleinert, Phys.\ Rev.\ D {\bf 57}, 2264 (1998)
(www.physik.fu-berlin.de/\~{}kleinert/257);
 Phys.\ Lett.\ B {\bf 434}, 74 (1998).


\end{thebibliography}
\end{document}